\documentclass[12pt]{article}
\usepackage{graphicx}
\begin{document}

\def\mn{{\medskip\par\noindent}}
\def\bn{{\bigskip\par\noindent}}
\def\sn{{\smallskip\par\noindent}}
\def\sig{{\sigma}}
\def\olig{{oligonucleotide}}  \def\oligs{{\olig s}}
\def\oligm{{oligomer}}  \def\oligms{{\oligm s}}
\def\kmer{{$k$-mer}} \def\kmers{{$k$-mers}}
\def\dist{{distribution}}  \def\dists{{\dist s}}
%%%% code for the organisms
\def\Atha{{\it Arabidopsis thaliana}}  \def\atha{{\it A. thaliana}}
\def\Aaeo{{\it Aquifex aeolicus}}  \def\aaeo{{\it A. aeolicus}}
\def\Aful{{\it Archaeoglobus fulgidus}}    \def\aful{{\it A. fulgidus}}  
\def\Aper{{\it Aeropyrum pernix}}  \def\aper{{\it A. pernix}} 
\def\Atum{{\it Agrobacterium tumefaciens}}  \def\atum{{\it A. tumefaciens}} 
\def\Bbur{{\it Borrelia burgdorferi}}  \def\bbur{{\it B. burgdorferi}}  
\def\Bhal{{\it Bacillus halodurans}} \def\bhal{{\it B. halodurans}} 
\def\Bmel{{\it Brucella melitensis}}   \def\Bmel{{\it B. melitensis}}   
\def\Bsub{{\it Bacillus subtilis}} \def\bsub{{\it B. subtilis}} 
\def\Buch{{\it Buchnera sp. APS}}   \def\buch{{\it B. sp.}}   
\def\Cace{{\it Clostridium acetobutylicum}}\def\cace{{\it C. acetobutylicum}}
\def\Cele{{\it Caenorhabditis elegans}} \def\cele{{\it C. elegans}}  %nematode
\def\Cglu{{\it Corynebacterium glutamicum}} \def\cglu{{\it C. glutamicum}}  
\def\Cjej{{\it Campylobacter jejuni}} \def\cjej{{\it C. jejuni}} 
\def\Clim{{\it Chlorobium limicola}} \def\clim{{\it Ch. limicola}} 
\def\Cmur{{\it Chlamydia muridarum}}  \def\cmur{{\it C. muridarum}}  
\def\Cper{{\it Clostridium perfringens}}  \def\cper{{\it C. perfringens}}   
\def\Cpne{{\it Chlamydia pneumoniae}}  \def\cpne{{\it Ch. pneumoniae}}  
\def\Ctra{{\it Chlamydia trachomatis}} \def\ctra{{\it Ch. trachomatis}} 
\def\Cvib{{\it Caulobacter vibrioides}}  \def\cvib{{\it C. vibrioides}} 
\def\Drad{{\it Deinococcus radiopugans}} \def\drad{{\it D. radiopugans}} 
\def\Dmel{{\it Drosophila melanogaster}} \def\dmel{{\it D. melanogaster}} 
\def\Ecol{{\it Escherichia coli}} \def\ecol{{\it E. coli}} 
\def\Fhep{{\it Flavobacterium heparinum}} \def\fhep{{\it F. heparinum}} 
\def\Fnuc{{\it Fusobacterium nucleatum}} \def\fnuc{{\it F. nucleatum}} 
\def\Gmax{{\it Glycine max}}  \def\gmax{{\it G. max}}    % Soybean
\def\Hasp{{\it Halobacterium sp.}}  \def\hasp{{\it H. sp.}} 
\def\Haur{{\it Herpetosiphon aurantiacus}} \def\haur{{\it H. aurantiacus}} 
\def\Hinf{{\it Haemophilus influenzae}} \def\hinf{{\it H. influenzae}} 
\def\Hpyl{{\it Helicbacter pylori}}  \def\hpyl{{\it H. pylori}}  
\def\Hsap{{\it Homo sapiens}} \def\hsap{{\it H. sapiens}} 
\def\Hvol{{\it Halobacterium volcanii}}  \def\hvol{{\it H. volcanii}}  
\def\Llac{{\it Lactococcus lactis}}  \def\llac{{\it L. lactis}}   
\def\Lmon{{\it Listeria monocytogenes}} \def\lmon{{\it L. monocytogenes}} 
\def\Mfer{{\it Methanothermus fervidus}} \def\mfer{{\it M. fervidus}} 
\def\Mgen{{\it Mycoplasma genitalium}} \def\mgen{{\it M. genitalium}} 
\def\Mjan{{\it Methanococcus janaschii}} \def\mjan{{\it M. janaschii}} 
\def\Mlep{{\it Mycobacterium leprae}} \def\mlep{{\it M. leprae}} 
\def\Mlot{{\it Mesorhizobium loti}}  \def\mlot{{\it M. loti}}  
\def\Mmus{{\it Mus musculus}} \def\mmus{{\it M. musculus}} % House mouse
\def\Mpne{{\it Mycoplasma pneumoniae}} \def\mpne{{\it M. pneumoniae}} 
\def\Mthe{{\it Methanobacterium thermoautotrophicum}} 
                     \def\mthe{{\it M. thermoautotrophicum}}  
\def\Mtub{{\it Mycobacterium tuberculosis}} \def\mtub{{\it M. tuberculosis}} 
\def\Nmen{{\it Neisseria meningitidis}} \def\nmen{{\it N. meningitidis}} 
\def\Neis{{\it Neisseria}}
\def\Nost{{\it Nostoc sp.}} \def\nost{{\it N. sp.}} % sp. PCC 7120 
\def\Paby{{\it Pyrococcus abyssi}}  \def\paby{{\it P. abyssi}}   
\def\Paero{{\it Pyrobaculum aerophilum}}  \def\paero{{\it P. aerophilum}}   
\def\Paeru{{\it Pseudomonas aeruginosa}} \def\paeru{{\it P. aeruginosa}} 
\def\Pfur{{\it Pyrococcus furiosus}}  \def\pfur{{\it P. furiosus}}   
\def\Phor{{\it Pyrococcus horikoshii}} \def\phor{{\it P. horikoshii}} 
\def\Pmul{{\it Pasteurella multocida}} \def\pmul{{\it P. multocida}} 
\def\Rcon{{\it Rickettsia conorii}} \def\rcon{{\it R. conorii}} 
\def\Rpro{{\it Rickettsia prowazekii}} \def\rpro{{\it R. prowazekii}} 
\def\Rsol{{\it Ralstonia solanacearum}} \def\rsol{{\it R. solanacearum}} 
\def\Saur{{\it Staphylococcus aureus}} \def\saur{{\it S. aureus}} 
\def\Sent{{\it Salmonella enterica}} \def\sent{{\it S. enterica}} 
\def\Scer{{\it Saccharomyces cerevisiae}} 
                     \def\scer{{\it S. cerevisiae}} % c./yeast
\def\Smel{{\it Sinorhizobium meliloti}}  \def\smel{{\it S. meliloti}}  
\def\Spne{{\it Streptococcus pneumoniae}} \def\spne{{\it S. pneumoniae}}  
\def\Spyo{{\it Streptococcus pyogenes}} \def\spyo{{\it S. pyogenes}} 
\def\Ssol{{\it Sulfolobus solfataricus}} \def\ssol{{\it S. solfataricus}} 
\def\Stok{{\it Sulfolobus tokodaii}} \def\stok{{\it S. tokodaii}}  
\def\Stub{{\it Solanum tuberosum}} \def\stub{{\it S. tuberosum}} % white potato
\def\Styp{{\it Salmonella typhimurium LT2}} \def\styp{{\it S. typhimurium}} 
\def\Syne{{\it Synechococcus sp.}} \def\syne{{\it S. sp.}}
\def\Taci{{\it Thermoplasma acidophilum}}  \def\taci{{\it T. acidophilum}}   
\def\Tmar{{\it Thermotoga maritima}} \def\tmar{{\it T. maritima}} 
\def\Tpal{{\it Treponema pallidum}} \def\tpal{{\it T. pallidum}} 
\def\Tten{{\it Thermoprotues tenax}}  \def\tten{{\it T. tenax}}  
\def\Tvol{{\it Thermoplasma volcanium}} \def\tvol{{\it T. volcanium}}  
\def\Uure{{\it Ureaplasma urealyticum}} \def\uure{{\it U. urealyticum}} 
\def\Vcho{{\it Vibrio cholerae}} \def\vcho{{\it V. cholerae}} 
\def\Xfas{{\it Xylella fastidiosa}} \def\xfas{{\it X. fastidiosa}} 
\def\Ypes{{\it Yersinia pestis}}  \def\ypes{{\it Y. pestis}}  
%%%%%%%%% End of genome names %%%%%%%%%%%%%%
\baselineskip=11pt
\twocolumn[\hsize\textwidth\columnwidth\hsize\csname %
@twocolumnfalse\endcsname
%]  Start one column

\begin{flushleft}
\begin{huge}
%{\bf\textsf{Microbial genomes are large systems with small-system 
%statistics - 
{\bf\textsf{Evidence for growth of microbial genomes by short 
segmental duplications}}
\end{huge}

\mn
{\bf\textsf{Li-Ching Hsieh$^*$, Liaofu Luo$^\|$ and 
\renewcommand{\thefootnote}{\fnsymbol{footnote}}
H.C. Lee$^{*}$\footnote[2]{\sl Correspondence and requests for material should 
be addressed to HCL at Physics, NCU; hclee@phy.ncu.edu.tw.}$^{\S}$}
\renewcommand{\thefootnote}{\arabic{footnote}}
}

\sn \baselineskip=7pt
{\scriptsize\textsf{$^*$Department of Physics and 
$^\dagger$Department of Life Science,
National Central University, Chungli, Taiwan 320;
$^\|$Department of Physics, Inner Mongolia University, Hohot, China;
$^\S$Centre de Recherches Math\'{e}matiques, 
Universit\'{e} de Montr\'{e}al, Montr\'{e}al, QC, Canada}
}

\end{flushleft}

] %end one column
\baselineskip=11pt
\begin{small}
\baselineskip=11pt
%%%%%% .txt file accompanying the manuscript gro_nat.doc %%%%%%%%%%%
%%%%%%  Started in 2002 October 3.  %%%%%%%%%%%%%%%%%%%%%%%%%%%%%%%%
%%%%%%%%%%%%%%Text for the section headed by the Nature style tag <p>
%%%%%%%%%%%%%%------------------------------------------------------

\noindent
{\sf We show that textual analysis of microbial genomes reveal telling
footprints of the early evolution of the genomes.  The frequencies of
word occurrence of random DNA sequences considered as texts in their
four nucleotides are expected to obey Poisson distributions.  It is
noticed that for words less than nine letters the average width of the
distributions for complete microbial genomes is many times that of a
Poisson distribution.  We interpret this phenomenon as follows: the
genome is a large system that possesses the statistical
characteristics of a much smaller ``random'' system, and certain
textual statistical properties of genomes we now see are remnants of
those of their ancestral genomes, which were much shorter than the
genomes are now.  This interpretation suggests a simple biologically
plausible model for the growth of genomes: the genome first grows
randomly to an initial length of approximately one thousand
nucleotides (1k nt), or about one thousandth of its final length,
thereafter mainly grows by random segmental duplication.  We show that
using duplicated segments averaging around 25 nt, the model sequences
generated possess statistical properties characteristic of present day
genomes.  Both the initial length and the duplicated segment length
support an RNA world at the time duplication began.  Random segmental
duplication would greatly enhance the ability of a genome to use its
hard-to-acquire codes repeatedly, and a genome that practiced it would
have evolved enormously faster than those that did not.}

\mn The genome is a highly complex network of embedded codes generated
in a very long process of evolution co-driven by chance mutations and
misreplications on the one hand and natural selection on the other.
The fact that both processes are stochastic makes it that much harder
to uncover what the the earliest genome looked like when life first
arose.  Adding the extreme diversity of organisms to the complexity of
each genome would seemingly render the task of unmasking the early
genome even more daunting.  It is therefore significant when a large
set of diverse and complex genomes share an unexpected common or
universal property.  Here we report one kind of universality in the
textual property of the genomes that allows us to deduce a mode of
growth which could be common to all early genomes.

\mn
{\bf{\textsf Frequency of occurrence of \oligs\ in microbial genomes}}
\sn
It is a general rule of statistics that very large systems have
sharply defined average properties.  When apples are randomly dropped
into barrels, the distribution of apples in the barrels is governed by
the Poisson distribution.  If 1,024 apples were dropped into
sixty-four barrels, in 95 of 100 cases, each barrel will have between
eight and twenty-four apples.  In comparison, if 1 million apples were
dropped into sixty-four barrels, in 95 of 100 cases, each barrel will
have between 15,875 and 15,375 apples.  There is a less than one in
$10^{830}$ ($10^{980}$, respectively) chance that one barrel would get
as many (few) as twenty-four (eight) thousand apples.

\mn
Microbial genomes are seemingly random systems when viewed as
texts of the four nucleotides represented by A, C, G and T.  To count
the number of times each of the sixty-four trinucleotides, or 3-mers,
occur in a genome-as-text is similar to counting apples in barrels.
The genome of the bacterium \Tpal, the causative agent of syphilis is
about 1M base pairs long and has almost even base composition
\cite{Tpal98}.  In an astonishing departure from what is expected of a
system of its size, the genome has six 3-mers (CGC, GCG, AAA, TTT,
GCA, TGC) occurring more than 24,000 times per 1M nt and two (CTA,
TAG) less than 8,000 times.  Scrambling the genome sequence thoroughly
restores it to a random sequence obeying Poisson distribution and the
large-system rule.

\mn \tpal\ is not exceptional in disobeying the
large-system rule.  For the fourteen complete microbial genome
sequences with approximately even base composition (see Methods), the
observed standard deviation (s.d.) of the distribution of the
frequency of occurrence (hereafter, simply distribution) of 3-mers per
1M nt is 4,080$\pm$630 around the mean of 15,625.  This is about 32
times the s.d. of a Poisson distribution typifying a random sequence 
with the same mean.

\mn Nor is the 3-mer exceptional in the \kmer-statistics of genomic
sequences.  In Table~\ref{t:deviation}, column 3 gives the average
s.d. of the distribution of \kmers\ per 1M nt, $k$ = 2 to 10, for the
fourteen genomic sequences and the s.d. of the average
(number given after the $\pm$ sign) and column 4 gives the
s.d. for a Poisson distribution (that describes a random sequence)
with mean value 10$^6$/4$^k$.  The s.d.'s of the genomic and random
sequence have about the same magnitude when $k$ is equal to or greater 
than 10 (not shown in the Table).  But with decreasing values of $k$
the Poisson s.d. increases as $2^{-k}$ whereas the genomic
s.d. increases at a much higher rate, such that for $k\le 8$ 
the Poisson s.d. is many times less than the genomic s.d. 
Moreover, the uncertainty in the genomic s.d. is typically much 
smaller than the difference between the genomic and Poisson s.d.'s. 
For example, at $k$=2 ($k$=6) the genomic s.d. is 40$\pm$8 
(9.0$\pm$1.3) times greater than the Poisson s.d.  Thus the genomic
distribution differs from the Poisson distribution in a universal
fashion, and in this sense we shall speak of a universal genome.

\begin{table}
\begin{footnotesize}
\caption{\label{t:deviation}
\scriptsize \baselineskip=9pt
\sf Standard deviation of \kmer\ distributions: for
the genome of \tpal; averaged over 14 microbial genomes 
with unbiased base composition; of a random sequence with 
Poisson distribution; of the model genome described in text.
In the third column, the number after the $\pm$ sign gives 
the s.d. associated with the average s.d. 
The last column is the length ($L_{eff}$) of a random sequence with the 
genomic ratio of mean count to s.d..}
\vspace{4pt}
%\begin{center}
\hspace{-0.4cm}
\vbox{
\begin{tabular}[h]{ccccc|c}
%\begin{tabular}[t]{ccccc|c}
\hline
$k$&{\it T. pal}&Genomic&Poisson&Present&$L_{eff}$\\ 
&&average&&model& (in k nt)\\
\hline
 2 & 8227 & 10580$\pm2040$ & 250 & 8207 & .65$\pm.35$\\
 3 & 3977 & 4080$\pm630$ & 125 & 3415 & 1.0$\pm0.3$\\
 4 & 1384 & 1490$\pm210$ & 62.5 & 1202 & 1.9$\pm0.5$\\
 5 & 434  & 469$\pm66$  & 31.2 & 402 & 4.7$\pm1.3$\\
 6 & 129  & 141$\pm21$  & 15.6 & 134 & 13$\pm4$\\
 7 & 37.5  & 41.9$\pm6.7$  & 7.8 & 45.3 & 37$\pm12$\\
 8 & 11.0  & 12.4$\pm2.3$  & 3.9 & 15.9 & 110$\pm40$\\
 9 & 3.4  & 3.84$\pm$0.84  & 1.9 & 5.9 & 300$\pm130$\\
 10 & 1.3  & 1.33$\pm$0.34  & 1.0 & 2.3 & 640$\pm300$\\
\hline
\end{tabular}
%\end{center}
}
\vspace{-0.6cm}
\end{footnotesize}
\end{table}

\mn
{\bf{\textsf Microbial genomes are large systems with small-system 
statistics}}
\sn
The universal genome has the statistical property
of a random sequence much smaller than itself.  To see this, we  
define the effective random-sequence length $L_{eff}$  of the 
universal genome as the length  
of a random sequence that has a \kmer\ \dist\ with 
a mean to s.d. ratio equal to that of the corresponding genomic 
ratio $r$.  Then $L_{eff}$=$4^k r^2$, and its values for the various $k$'s  
are given in the last column of Table~\ref{t:deviation}.    One
notices that the $L_{eff}$ of 
the universal genome is very short for
the smaller $k$'s - of the order of 1k nt for $k$$\le$3 -  
and grows with $k$.  When $k$=10, it is essentially
the same length as the real genome.

\mn
A signature of the universal genome is that compared to a random 
sequence, the former has very large numbers of both overrepresented 
and underrepresented \oligs.  As a typical representative of the 
universal genome, the genome of \ecol\ \cite{ecol} 
has 500 and 510 6-mers whose 
frequency of occurrences are greater than 400 and 100 per 1M nt, 
respectively, while a random sequence has none in either category.   
There are many known examples of individual \olig\ that exhibit 
extreme relative abundance. 
For dinucleotides this was noted to be common and has genome-wide 
consistency \cite{Karlin95}; tetra- and
hexapalindromes are almost always underrepresented in bacteriophages
and are underrepresented systematically in bacteria where 4-cutting
and/or 6-cutting restriction enzymes are common \cite{Karlin92}; an
8-mer that appears as Chi sites, hotspots of homologous recombination,
is highly overrepresented in \ecol\ \cite{Colbert98}; in the human
pathogens \Hinf\ \cite{Smith95,Karlin96} and \Neis\ \cite{Smith99} 
there are 9- and 10-mers functioning as uptake signal sequences 
that are vastly overrepresented.   The causes for these extreme cases 
are generally not known and, with the exception of the dinucleotides, 
these individual cases do not much affect the statistical properties of 
the genome.  

\mn What caused a genome to have statistical characteristics so starkly
distinct from those of a random sequence?  Natural selection suggests
itself as a prime explanatory candidate.  For instance, the 64
frequencies of codons, 3-mers used by the genome to code proteins in
genes, exhibit very wide distributions.  But natural selection by
itself does not directly cause any change in a genome.  Such changes
are caused by mutation and other mechanisms, all believed to occur at
random.  Natural selection may account for what changes come to pass;
if, however, such changes always tend to promote or retain a
randomness that exhibits Poisson distribution, then the ability of
natural selection to push the genome very far in a non-Poisson
direction would seem to have its limits.

\mn
{\bf{\textsf Model for early genome growth}}
\sn
Here we propose a biologically plausible model for the growth and
evolution of a universal genome that can generate the observed
statistical characteristics of genomic sequences.   The model is very
simple and consists of two phases.  In the first phase the genome
initially grows to a random sequence whose size is much smaller than
the final size of the genome.  In the second phase the genome grows by
random duplications modulated by random single mutations.  In this
work a snapshot is taken of the model genome shortly after it reaches
a length of 1M nt.  The key in the model is growth by duplication; 
it is most straightforward way for the universal genome to become 
what it appears to be: a large system that exhibits small-system 
statistical characteristics. 

\mn
We found it comparatively easy to generate a sequence that could
faithfully reproduce the genomic \kmer\ distribution of a particular
$k$ but not those of other $k$'s.  Typically such a sequence had an
excessively rigid effective random-sequence length and, consequently,
a distribution too narrow (broad) for smaller (greater) $k$'s.
Several such examples are given in the Methods.  Generating a sequence
that would emulate a real genome was a much more exacting task.

\begin{figure} [t!]
\begin{center}
\includegraphics[width=3.5in,height=2.5in]{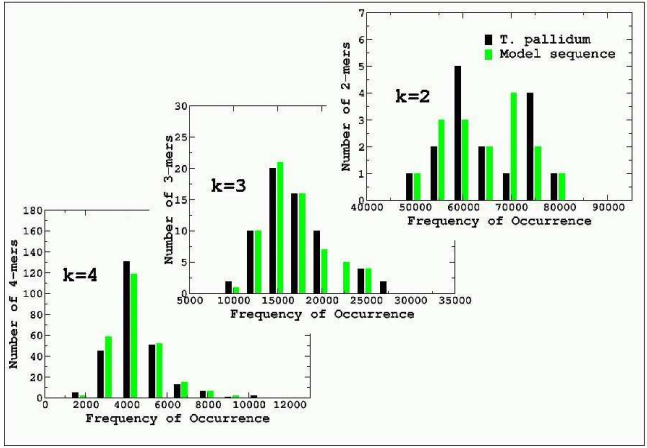}
\end{center}
\vspace{-15pt}
\caption{\label{Tpal_his_2-4} \scriptsize\baselineskip=9pt\sf 
Histograms of \kmer\ distributions of genome of 
{\it T. pal.} (black) and model sequence (gray/green), $k$=2 to 4.
Abscissa indicate intervals of frequency of occurrence of \kmers; 
ordinates give the number of \kmers\ falling within a given 
interval of frequency of occurrence.  In each case the histogram of 
the distributions for a random sequence would be represented by 
a single tower located at the mean frequency.} 
\vspace{-10pt}
\end{figure}

\mn
{\bf{\textsf Result}}
\sn 
After extensive experimentation, it was found that sequences
having the statistical characteristics sought after could be
generated from an initial random sequence approximately 1k nt long
($L_0$) which was then grown to 1M nt by random duplication of segments
of length ($\bar{l}$) averaging 25 nt with a spread ($\Delta_l$) of 
approximately 11 nt (see Methods for detail).

\mn
The s.d. of the \kmer\ distribution of a good model
sequence are given in column five of Table~\ref{t:deviation}.  They
agree quite well with the observed genomic values in columns two and
three although their $k$-dependence is slightly too strong.
Histograms in Fig.~\ref{Tpal_his_2-4} show comparisons between the \kmer\
distributions for $k$=2, 3 and 4 of the genome of \tpal\ (black)
and those of the model sequence (green/gray).  
In all three cases, the histogram for a random sequence would be
represented by a single tower located at the mean frequency.  For
$k$=2 and to a lesser extent $k$=3, the histograms for both genomic
and model sequences display large fluctuations.  The model sequence is
not expected to exactly reproduce the counts of the genomic sequence.
Indeed, generated stochastically, another (good) model sequence
would give distributions indistinguishable from those shown in
Fig.~\ref{Tpal_Repk5-7} but something rather different than those shown
in the $k$=2 and 3 panels of Fig.~\ref{Tpal_his_2-4}.  In any case, all
model sequences would show patterns of fluctuation similar to those
exhibited by the genomic sequence and have s.d.'s similar
to those given in column 5 of Table~\ref{t:deviation}.
Fig.~\ref{Tpal_Repk5-7} shows comparisons for $k$=5 to 9.  
The panel at the top-left corner compares the
6-mer distribution from \tpal\ with that of a random sequence obtained
by scrambling the \tpal\ genome.  The strong agreement between the
microbial genome and the model sequence contrasts sharply with the
glaring differences between the genome and the random sequence.

\begin{figure} [t!]
\begin{center}
\vspace{5pt}
\includegraphics[width=3.5in,height=2.5in]{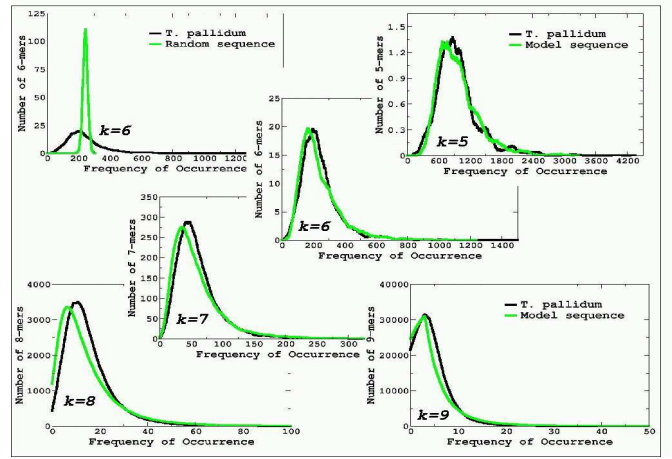}
\end{center}
\vspace{-15pt}
\caption{\label{Tpal_Repk5-7} \scriptsize\baselineskip=9pt\sf 
Comparison of \kmer\ distributions, $k$= 5 to 9.  Abscissa give the 
frequency of occurrence of a \kmer; ordinates give the number 
of \kmers\ having a given frequency of occurrence.  Black: the 
distribution from the genome of \tpal; Gray (or green): the distribution from 
the simulated model sequence.  Top-left panel:  {\it T. pal.} and random 
sequence, $k$=6.  Other panels: {\it T. pal.} and model sequence.} 
\vspace{-10pt}
\end{figure}

\mn The model sequence is parameter-sensitive: If $L_0$ was much 
longer than 1k nt no good model sequence could be found (this is 
expected because $L_0$ cannot be much longer than the shortest 
$L_{eff}$ in Table~\ref{t:deviation}); if either $\bar{l}$ 
or $\Delta_l$ was was changed by more than 10\% from their 
optimal values of 25 nt and 11 nt respectively the agreement between 
the genomic and model sequences would worsen noticeably (see Methods).  
No mutations were imposed on the model sequence whose properties are
shown here; twenty thousand mutation fixations reduces the 
s.d. of the \kmer\
distributions of the model sequence by 4\% (for $k$=2) to 10\%
($k$=10) but under casual inspection the model sequence - with or
without mutation - has the appearance of a random sequence.  Results
showing the model reproducing the \kmer\ \dists\ of microbial genomes
with highly biased compositions will be presented elsewhere.

\mn 
In bacterial genomes, typically about 12\% of genes represent recent
duplication events - 12\% in \tpal\ \cite{Tpal98}, 11.2\% in \hinf\
\cite{arabidopsis} and 12.8\% in \vcho\ \cite{Vcho02}.  Our model
sequence as presented here does not yet fully explain the pattern of
all such duplications, many of which would involve segments up to
several k nt long.  Work is under way to extend the model to account for
the genomic pattern of repeat sequences of all lengths.

\mn
{\bf{\textsf Discussion}}
\sn
We mention some biological and evolutionary implications assuming our
model does capture the essence of the early growth mechanism of
microbial genomes and, by extension, perhaps of all genomes.  Setting
the initial length of our model universal genome before it began the 
growth by duplication process to about 1k nt but not much longer 
(as required by observed data) necessarily implies that the
universal genome began its life in an RNA world
\cite{Gilbert86,Darnell86} in which there were no proteins and RNAs
had the dual roles of genotype and phenotype (see \cite{Klyce02} for a
review).  This view of the origin of life
\cite{Woese67,Crick68,Orgel68} gained much credence when RNA was
discovered to exhibit self-splicing and enzymatic activities
\cite{Cech81,Takada83}.  Some RNA enzymes, or ribozymes, are very
small; the hammerhead ribozyme is only 31 to 42 nt long
\cite{Forster87} and the hairpin ribozyme is only 50 nt long
\cite{Hampel89}.  Thus the 1k nt initial universal genome was certainly 
of sufficient size to possess a machinery for sustained evolution and
duplication.  Our model does not address the origin of this initial
genome.  The likelihood of its being the evolutionary product of something
that arose spontaneously is enhanced by the succcessful isolation of
artificial ribozymes from pools of random RNA sequences
\cite{Ekland95}.  The average duplicated segment length of 25 nt is
very short compared to a present-day gene that codes for a protein,
but likely represents a good portion of the length of a typical
ribozyme encoded in the early universal genome.

\sn Shifting the burden away from natural selection onto segmental
duplication as the main force driving the universal
genomes so far in a non-Poisson direction implies a much higher
evolution rate than it might have been if natural selection were 
the only driving force.  The model suggests that
uneven codon usage was not the primary cause of the very broad
distribution of the 3-mer counts seen in the universal genomes.
Rather, the rise of codon was the consequence of an opportunistic
evolutionary adaptation to the already-wide 3-mer distribution that
had resulted from growth by duplication.  Similarly, many - but not all
- of the highly under- or overrepresented \oligs\ we see now must have
been recruited for their respective biological functions after they
already had (the suitable beginnings of) biased frequencies of
occurrence.

\sn That some statistical characteristics of a present day genome are
determined by the charateristics of the genome when it first began to
grow by duplication means that we should be able to learn something
about such early genomes, and each such ancestral genome should be
common to a group of present day genomes that are phylogentically
close.  Detailed analyses made along this line of reasoning may bring 
us a step nearer in understanding the univeral ancestor \cite{Woese98}.

\sn Being a natural way to repeatedly utilize hard-to-come-by codes,  
growth by duplication is in itself a brilliant strategy and 
must have increased the rates of evolution and species
diversion enormously.   The continuity of this strategy after the 
rise of codons and proteins is bundantly in evidence.   In higher 
organisms a large number of repeat sequences with lengths ranging
from 1 base to many kilobases are believed to have resulted from 
at least five modes of duplication \cite{Lander01,Venter01}.  
This strategy should provide part of the answer to the questions 
\cite{Meyer03}: how have genes been duplicated at the high rate of
about 1\% per gene per million years \cite{Lynch00}? and why are there
so many duplicate genes in all life forms \cite{Maynard98,Otto01}?
The fact that duplicate genes (after they have diverged) contribute to
genetic robustness by protecting the genome against harmful mutations
\cite{Gu03} is likely not what caused the proliferation of duplicate
genes, but is rather another example of an adaptation to an existing 
situation by natural selection for a beneficial function.

\end{small}

\begin{footnotesize}  
\baselineskip=10pt
\mn
{\small\bf{\textsf Methods}}
\sn
{\bf The fourteen microbial genome sequences} (length ($L$) in M nt and
G+C probability ($p$) in brackets) \ecol\ K12 (4.64, .50), \ecol\
0157 (5.52, .50), \mthe\ (1.75, .50), \aful\ (2.18, .49), \tpal\
(1.14, .53), \xfas\ (2.67, 0.53), \vcho\ chromosomes I (2.96, .48) and
II (1.07, .47), \Syne\ (3.57, .48), \nmen\ serogroup B strain MC58
(1.57, .52), \ypes\ (4.65, .48), \styp\ (4.86, .52), \sent\ (4.81,
.52) and \paero\ (2.22, .51) are obtained from the GenBank
\cite{GenBank}.  Counting of \kmers\ is done by reading through 
a $k$-base wide window that is slid around the (circular) genome once. 
Counts are normalized to per 1M nt and bias in base composition is 
corrected for by dividing the actual counts by the factor 
$L 2^k p^{n} (1-p)^{k-n}$, where $n$ is the total number 
of G's and C's in each \kmer.

\sn
{\bf Generation of model sequence}.  A random sequence of length $L_0$
is first generated.  Thereafter the sequence is altered by single
mutations (replacements only) and duplications, with a fixed average
mutation to duplication event ratio.  In duplication events, a segment
of length $l$, chosen according to the Erlang probability density
function $f(l)=1/(\sig m!) (l/\sig)^m e^{-l/\sig}$, is copied from one
site and pasted onto another site, both randomly selected.  In the
above $m$ is an integer and $\sig$ is a length scale in bases.  The
function gives a mean duplicated segment length $\bar{l}=(m+1)\sig$
with s.d. $\Delta_l=(m+1)^{1/2}\sig$.  The values $m=0$
to 8 and selected values for $\sig$ from 3 to 15,000 were used.  The
model sequence compared with genomic sequences in the Figures 1 and 2
and in Table 1 was generated with $L_0=1000$, $m=4$, $\sig=5$ and
without mutation events. Fine-tuning to find the best parameters was
not attempted.  The following are some examples that gave very good
distributions for specific \kmers\ but not generally; all were
generated with $L_0=1000$ and $m=0$: for 6-mer, $\sig=13,000\pm 2,000$
and on average 0.04$\sig$ mutations per duplication (these parameters
also work for genomes with biased base compositions) \cite{Hsieh02};
for 2-mer, $\sig=50$, no mutation; for 5-mer, $\sig=30$, no mutation;
for 9-mer, $\sig=15$, no mutation.

\sn
{\bf Presentation of data}. In Fig.~\ref{Tpal_Repk5-7} the curves
shown are the result of a small amount of forward and backward
averaging - to remove excessive fluctuations.  In
Fig.~\ref{Tpal_his_2-4} data bunching was used to produce the towers
shown.

%\noindent 
%{\footnotesize\bf{\textsf Acknowledgment}}
\mn
{\scriptsize{HCL thanks the National Science Council (ROC) for the
grant NSC 91-2119-M-008-012 and members of the Redfield Lab and the 
Otto Lab, Department of Zoology, University of British Columbia, 
for discussion and the Institute for Theoretical Physics, Chinese 
Academy of Science, Beijing, and the Center for Theoretical Biology, 
Beijing University, for hosting visits.}}  

%\sn

\end{footnotesize}

%\sn\rule{3.6in}{.1mm}
%\sn\hrulefill
%%%% First get rid of ``References'' heading
%%%% then change labeling style
\renewcommand{\refname}{}   
\makeatletter  
\renewcommand{\@biblabel}[1]{\hfill#1.} 
\makeatother
\vspace{-1.2cm}
\begin{scriptsize}

\end{scriptsize}

\end{document}